\documentclass[preprint2]{aastex}

\newcommand{\dm}{$\Delta m_{15}$(B)}
\newcommand{\RSi}{$\cal R$(SiII)}
\newcommand{\vSiiX}{$v_{10}$(SiII)}
\newcommand{\kms}{km\,s$^{-1}$}
\newcommand{\kmsd}{km\,s$^{-1}$\,d$^{-1}$}

\shorttitle{Diversity of SN~Ia }
\shortauthors{Benetti et al.}

\received{2004 October 29}
\begin{document}


\title{The diversity of Type Ia Supernovae: evidence for systematics?}


\author{S. Benetti\altaffilmark{1}; E. Cappellaro\altaffilmark{2};
P. A. Mazzali\altaffilmark{3,4}; M. Turatto\altaffilmark{1};
G. Altavilla\altaffilmark{5}; F. Bufano\altaffilmark{1};
N. Elias-Rosa\altaffilmark{1};
R. Kotak\altaffilmark{6}; G. Pignata\altaffilmark{7};
M. Salvo\altaffilmark{8}; V. Stanishev\altaffilmark{9} }

\altaffiltext{1}{INAF-Osservatorio Astronomico, vicolo dell'Osservatorio 5, Padova, Italy}

\altaffiltext{2}{INAF-Osservatorio Astronomico di Capodimonte, Napoli, Italy}

\altaffiltext{3}{INAF-Osservatorio Astronomico, Trieste, Italy}

\altaffiltext{4}{Max-Planck-Institut f\"ur Astrophysik, Garching, Germany }

\altaffiltext{5}{Department of Astronomy, U. Barcelona, Barcelona, Spain}

\altaffiltext{6}{Blackett Laboratory-Imperial College, London, England}

\altaffiltext{7}{European Southern Observatory, Garching, Germany}

\altaffiltext{8}{Australian National University-Mt.Stromlo Observatory, Australia}

\altaffiltext{9}{Stockholm University, Stockholm, Sweden}

%




\begin{abstract}
The photometric and spectroscopic properties of 26 well observed Type
Ia Supernovae (SNe~Ia) were analyzed with the aim to explore SN~Ia
diversity.  The sample includes (Branch-)normal SNe as well as extreme
events like SNe~1991T and 1991bg, while the truly peculiar SN~Ia,
SN~2000cx \citep{li01} and SN~2002cx \citep{li03} are not included in
our sample.  A statistical treatment reveals the existence of three
different groups. The first group (FAINT) consists of faint SNe~Ia
similar to SN~1991bg, with low expansion velocities and rapid
evolution of SiII velocity. A second group consists of ``normal''
SNe~Ia, also with high temporal velocity gradient (HVG), but with
brighter mean absolute magnitude $<M_B>=-19.3$ and higher expansion
velocities than the FAINT SNe.  The third group includes both
``normal'' and SN~1991T-like SNe~Ia: these SNe populate a narrow strip
in the SiII velocity evolution plot, with a low velocity gradient
(LVG), but have absolute magnitudes similar to HVGs.  While the FAINT
and HVG SNe~Ia together seem to define a relation between \RSi\ and
\dm, the LVG ones either do not conform with that relation or define a
new, looser one. The \RSi\ pre-maximum evolution of HVGs is strikingly
different from that of LVGs. The impact of this evidence on the
understanding of SN~Ia diversity, in terms of explosion mechanisms,
degree of ejecta mixing, and ejecta-CSM interaction, is discussed.
\end{abstract}



\section{Introduction}\label{intro}

Given the role of Type Ia Supernovae as distance indicators for cosmology and
main producers of heavy elements in the Universe, understanding the physics
of their explosions and how it influences observables is one of the most
fundamental issues in modern Astrophysics.  One of the keys to penetrate into 
the secrets of SN~Ia physics is to explore the origin of their diversity.

During the last decade a new paradigm for SNe~Ia was developed. In
particular, a correlation between the peak luminosity and the shape of
the early light curve was found, with brighter objects having a slower
rate of decline than dimmer ones (Phillips 1993; Phillips et al. 1999,
hereinafter P99). This is matched by a spectroscopic sequence, defined
by the ratio of the depth of two absorption features of SiII at 5972 and 6355
\AA\ (typically observed at 5800 and 6150 \AA, respectively)
\citep{nug95}. This ratio, $\cal R$(SiII), also correlates with the
absolute magnitude of SNe~Ia and, in turn, with the rate of
decline. Spectral modeling indicates that most of the spectral
differences are caused by variations in the effective temperatures
which, in the context of Chandrasekhar-mass explosions, can be
interpreted in terms of a variation in the mass of $^{56}$Ni produced
in the explosions.\\
Alternatively, \citet{gar04}, using synthetic spectra, 
tentatively suggest that in SN~1999by and, in general,
in SNe~Ia with \dm$>1.2$ the 5800 \AA\ feature is
mostly due to TiII transitions rather than to SiII.

Although a one-parameter description of SNe Ia has proved to be very
useful, it does not completely account for the observed diversity of
SNe~Ia, (e.g.  \citep{ben04,pig04}). \cite{hat00} showed that,
contrary to expectations, $\cal R$(SiII) correlates poorly with the
photospheric velocity deduced from the Si~II $\lambda 6355$
absorption. To account for this, they suggested that two or more
explosion mechanisms are required to explain SN~Ia variety.
Furthermore, some SNe~Ia with normal spectra (i.e. showing only lines
of typical ions) were noticed to show exceptionally high absorption
line blueshifts \citep{bra87}.  Finally, no correlation was found
between the blueshift of the Si~II 6355 absorption at the time of
maximum brightness and the decline rate parameter, \dm, in a small
sample of well-observed SNe Ia (see also Patat et al.  1996).

In this paper we further explore this issue, using detailed
observations of a large sample of SNe~Ia.

\section{Analysis}\label{}

The sample used in this work consists of 26 well-studied SNe Ia. It
includes (Branch-)normal SNe \citep{bra93a} as well as extreme events
like SNe 1991T and 1991bg. Photometric parameters such as \dm,
extinction and apparent magnitude at maximum are from P99 or
alternatively from \citet{alt04}. The spectral parameters have been
(re-)measured homogeneously on the available spectra, most of which
are published although unpublished material collected in the Asiago
Supernova Archive (ASA) by the Padua-Asiago SN group and the European
Supernova Collaboration was also used. In Table \ref{tab1} we
summarize the main photometric and spectroscopic parameters of the
SN~Ia sample, together with the morphological type (T) of the host
galaxy as given in RC3. The SNe are divided into three groups
according to the criteria discussed in Sect. \ref{exp}. In each group 
the SNe are arranged by decreasing $\dot{v}$ (see Sect. \ref{exp}).

\begin{figure*}
\figurenum{1}
\includegraphics[height=12cm, angle=270]{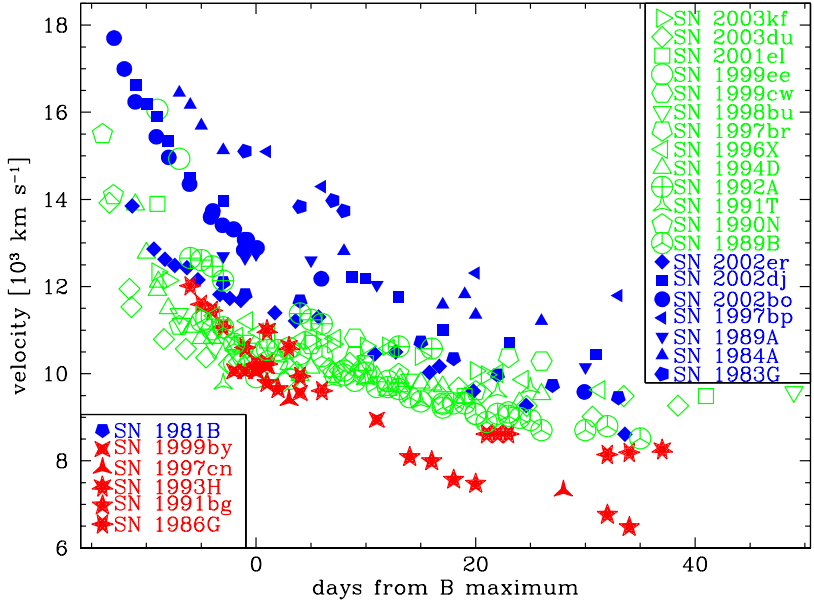}
\caption{Time evolution of the photospheric velocity derived from 
SiII $\lambda 6355$ for the SNe of Tab.\ \ref{tab1}. The filled symbols refer to HVG,
open symbols to LVG and starred to FAINT SN~Ia. }
\end{figure*}

\subsection{Expansion velocities from SiII} \label{exp}

The expansion velocity of the ejecta gives a direct indication of the
kinetic energy of the explosion. The blue-shift of the SiII $\lambda
6355$ absorption, the most prominent line in the photospheric phase
spectrum, traces the evolution of the expansion velocity of the ejecta
over the first 4-5 weeks past explosion. A plot of the time evolution
of the expansion velocity, v(Si) for the SNe~Ia of our sample (Figure
1) confirms the results of \citet{bra87}. At any given phase,
expansion velocities span a wide range of $\sim 4000$\,\kms, from the
very rapidly expanding ejecta of SNe~1983G and 1997bp to the low
velocity, SN~1991bg-like events. Between these limits both the
``Branch normal'' SNe~Ia and the luminous SN~1991T-like events are
found. A convenient parameter to distinguish different SNe~Ia is the
expansion velocity measured 10 days past maximum, \vSiiX.  As we
mentioned earlier, this parameter correlates poorly with \RSi\/
\citep{hat00} or with \dm\/ \citep{pat96}, hence it may give
interesting clues for origin of SN~Ia diversity.

Upon a more careful scrutiny, and thanks to the excellent coverage
offered by the new data, it can be noticed that the velocities show
not only a spread in value, but also significant differences in
velocity evolution.  In particular, a sub-group of very homogeneous
SNe~Ia, characterized by a shallow evolution of the expansion velocity
(open symbols in Fig. 1), populates a narrow strip in the post-maximum
velocity evolution diagram.  For a quantitative analysis we introduce
a new parameter, $\dot{v} = -\Delta$v/$\Delta$t, which is the average
daily rate of decrease of the expansion velocity (Tab. \ref{tab1}, col. 4). 
This is derived from least squares fits of the measurements taken between 
maximum and either the time the SiII feature disappears or the last 
available spectrum, whichever is earlier.

On average, the low (temporal) velocity gradient (LVG) SNe~Ia have a
velocity gradient $\dot{v} < 60-70$ \kmsd.  The SNe~Ia shown as filled
symbols have a larger $\dot{v} > 70$ \kmsd, reaching 110-125 \kmsd~
for SNe~2002bo, 1991bg, and 1983G (although the value for SN~1983G is
quite uncertain).

A similar dichotomy in velocity slopes between SNe~Ia can be found in the
velocities deduced from the SII $\lambda 5640$ line, this time at pre-maximum
phases (see Figure 11 of \citet{ben04}).

For an objective identification of possibly homogeneous groups, we
performed a hierarchical cluster analysis for the SN~Ia sample of
Table \ref{tab1}, considering both photometric and spectroscopic
parameters, but neglecting the dependence on galaxy morphological
type. Hierarchical Cluster Analysis \citep{and73} is an
exploratory data analysis tool which aims to identify relatively
homogeneous groups of events based on selected characteristics, using
an algorithm that starts with each case in a separate cluster and
combines clusters until only one is left.  Cluster analysis simply
discovers structures in data without explaining why they exist. The
choice of the number of clusters to be considered is somewhat
arbitrary although a criterion is the distance of the groups in the
parameter space.\\ 
Indeed, in the five dimensions space we found three well separated
clusters:

\begin{description} 

\item[a)] A first cluster (FAINT) consists of faint SNe~Ia, with
  $<M_B> = -17.2$ ($\sigma=0.6$), similar to SN 1991bg.  All these SNe
  have a high post-maximum luminosity decline rate, $<$\dm$>=1.83$
  ($\sigma=0.09$), and SiII line ratio, $<$\RSi$>$=0.58
  ($\sigma=0.05$). They have small expansion velocities,
  $<$\vSiiX$>$=9.2 ($\sigma=0.6$), and a large velocity gradient,
  $<\dot{v}> = 87$ ($\sigma=20$).

\item[b)] A second group consists of ``normal'' SNe~Ia with high
velocity gradient (HVG), $<\dot{v}>=97$ ($\sigma=16$). 
These SNe have average absolute magnitude $<M_B> = -19.3, (\sigma=0.3)$, 
\dm=$1.2$ ($\sigma=0.1$), and $<$\RSi$>$=0.20 ($\sigma=0.05$). 
They typically have high expansion velocities, $<$\vSiiX$>$=12.2 
($\sigma=1.1$).

\item[c)] A third group consists of SNe with a low velocity gradient (LVG):
$<\dot{v}> = 37$ ($\sigma=18$), but it includes both ``normal'' SNe~Ia and 
all the brightest SNe. Although its post-maximum decline rate is somewhat 
slower than that of the HVG SNe, $<$\dm$>$=1.1 ($\sigma=0.2$), its average 
luminosity is similar: $<M_B> = -19.2 (\sigma=0.3)$, and so is their 
$<$\RSi$>$=0.25 ($\sigma=0.07$).  
On average, the LVG SNe have lower and more homogeneous expansion velocities than the HVG SNe, 
$<$\vSiiX$>$=10.3 ($\sigma=0.3$).
\end{description}

For a few events, in particular SNe 1989B and 1992A, cluster membership is
uncertain. A small variation of the parameters, still within the errors, could
shift them from the LVG to the HVG group.

It is also interesting that on average the host galaxy morphological type is
progressively later as one moves from FAINT to HVG and to LVG. The average
values are $-1.4$, 0.6, and 2.5, respectively. The dispersion within each group
is however very large ($\sigma=3-4$). This is consistent with the finding that
bright SNe~Ia occur preferentially in late type galaxies, while faint SNe~Ia 
are more often found in early type galaxies \citep{alt04}.\\

\begin{table*}
\begin{flushleft}
\scriptsize
\caption{Observed parameters of SN~Ia sample, the host galaxies and references.
 \label{tab1}}
\begin{tabular}{crrrrrrl}
\tableline\tableline
SN  & \dm$^*$ & M$_B^{**}$ & $\dot{v}$ ~~~~~~& $v_{10}$(SiII)$^{***}$~&\RSi$_{max}^{****}$& T$_{(RC3)}$&References\\
    &         &            &[kms$^{-1}$d$^{-1}]$&[kms$^{-1}$1000$^{-1}$]&          &            & \\
\tableline
\multicolumn{6}{c}{LVG}\\
92A & $1.47\pm0.05$& -18.81& $67\pm7$&$10.83\pm0.15$ & $0.38\pm0.05$&-1.9&P99; ASA; K93 \\
89B & $1.34\pm0.07$& -18.87& $66\pm5$&$10.03\pm0.10$ & $0.29\pm0.05$& 3.0&P99; B90; W94 \\
03kf& $1.01\pm0.05$& -19.37& $50\pm5$&$10.63\pm0.10$ & $0.18\pm0.05$& 3.0&a \\
96X & $1.25\pm0.05$& -19.24& $46\pm5$&$10.68\pm0.15$ & $0.25\pm0.05$&-5.0&P99; S01 \\
99ee& $0.94\pm0.04$& -19.46& $42\pm5$&$ 9.90\pm0.10$ & $0.22\pm0.05$& 4.0&S02; H02 \\
90N & $1.08\pm0.05$& -19.23& $41\pm5$&$10.12\pm0.10$ & $0.21\pm0.05$& 3.8&P99; ASA; L91 \\
94D & $1.32\pm0.05$& -19.06& $39\pm5$&$ 9.89\pm0.10$ & $0.33\pm0.05$&-2.0&P99; P96 \\
03du& $1.06\pm0.06$& -18.93& $31\pm5$&$10.10\pm0.10$ & $0.22\pm0.03$& 8.0&b \\
01el& $1.15\pm0.04$& -18.71& $31\pm5$&$10.24\pm0.10$ & $0.30\pm0.07$& 5.9&K03; W03; c \\
97br& $1.04\pm0.15$& -19.62& $25\pm5$&$10.40\pm0.10$ & $<0.1^\dag$  & 7.0& ASA; L99 \\
99cw& $0.94\pm0.05$& -19.24& $22\pm7$&$10.79\pm0.15$ & $<0.31^{\dag\dag}$& 1.5& d \\
91T & $0.95\pm0.05$& -19.62& $11\pm5$&$ 9.87\pm0.10$ & $0.14\pm0.05$& 3.8&P99; P92; M95 \\
98bu& $1.04\pm0.05$& -19.12& $10\pm5$&$10.29\pm0.10$ & $0.21\pm0.05$& 2.0&P99; ASA; H00 \\
\tableline
\multicolumn{6}{c}{HVG}\\
83G & $1.37\pm0.10$& -18.62&$125\pm20$&$13.49\pm0.20$ & $0.28\pm0.06$&-2.2&H83; B85; T85; B91; B89; e \\
02bo& $1.17\pm0.05$& -19.42&$110\pm7$&$11.73\pm0.15$ & $0.17\pm0.05$& 1.0&B04 \\
97bp& $1.09\pm0.10$& -19.69&$106\pm7$&$13.87\pm0.20$ & $0.16\pm0.05$& \#&A04; ASA\\
02er& $1.33\pm0.04$& -19.45& $92\pm5$&$10.52\pm0.10$ & $0.23\pm0.05$& 1.0&P04; f \\
84A & $1.21\pm0.10$& -19.46& $92\pm10$&$12.64\pm0.15$& $0.23\pm0.05$& 1.0&B89; Ba89; W87; e \\
89A & $1.06\pm0.10$& -19.21& $90\pm10$&$12.13\pm0.15$& $0.22\pm0.07$& 4.1&B91; e\\
02dj& $1.12\pm0.05$& -19.05& $86\pm6$&$12.18\pm0.10$ & $0.17\pm0.05$&-5.0&g \\
81B & $1.11\pm0.07$& -19.21& $76\pm7$&$11.11\pm0.15$ & $0.16\pm0.05$& 4.5&P99; B83 \\
\tableline
\multicolumn{6}{c}{FAINT}\\
99by& $1.87\pm0.10$& -16.64&$110\pm10$&$ 9.06\pm0.10$ & $0.58\pm0.06$& 3.0&B99; V02; H01 \\
91bg& $1.93\pm0.10$& -16.81&$104\pm7$&$ 8.50\pm0.15$ & $0.62\pm0.05$&-4.7&P99; T96 \\
97cn& $1.86\pm0.10$& -16.95& $83\pm10$&$ 8.81\pm0.20$ & $0.63\pm0.06$&-5.0&T98 \\
93H & $1.70\pm0.10$& -18.20& $73\pm8$&$10.10\pm0.20$ & $0.52\pm0.05$& 1.9&P99; ASA; h \\
86G & $1.78\pm0.07$& -17.48& $64\pm5$&$ 9.36\pm0.15$ & $0.53\pm0.05$&-2.2&P99; P87; C92 \\

\tableline
\end{tabular}

* \small {reddening corrected according to P99}\\

** Cepheids distances taken from Altavilla et al. 2004
   (H$_0=72$km\,s$^{-1}$\,Mpc$^{-1}$) when available,
   otherwise relative distances to Virgo taken from \citet{kra82} and a Virgo
   distance of 15.3 Mpc \citep{fre01} or from Hubble flow\\

*** in units of 1000 \kms

**** measured at maximum light

\dag SiII $\lambda 5972$ not visible on the -3$^d$ spectrum; spectra  
close to maximum not available. 

\dag\dag measured on the earliest available spectrum($\phi=+5d$)

\# Irregular galaxy

a) Salvo et al. 2004, in preparation\\


b) Stanishev et al. 2004, in preparation\\

c) Mattila et al. 2004, in preparation\\

d) Bufano et al. 2004, in preparation\\

e) McDonald archive; 89A, 84A and 83G: determination of \dm~ from published data\\

f) Kotak et al. 2004, in preparation\\

g) Pignata et al. 2004, in preparation\\

h) CTIO Archive\\

\end{flushleft}
\end{table*}

\begin{figure}[t]
\figurenum{2}
\includegraphics[width=6.5cm, angle=-90]{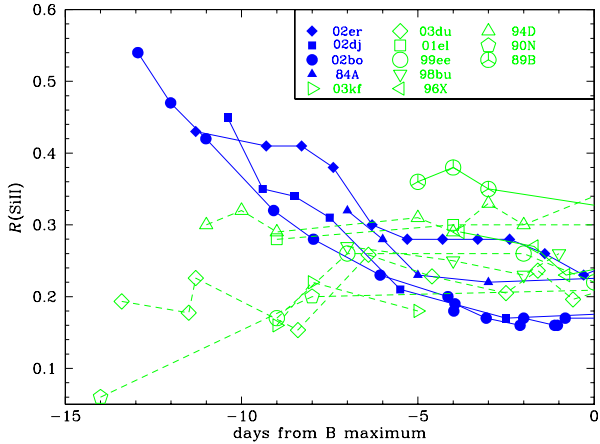}
\caption{The pre-maximum temporal evolution of the \RSi~ parameter for
our sample of SNe~Ia. Symbols are as in Fig. 1.}
\end{figure}

\subsection{Pre-maximum evolution of \RSi}

Figure 2 shows the pre-maximum evolution of \RSi\ for our sample of
SNe~Ia.  Interestingly, the HVG SNe~Ia for which very early
observations are available show a dramatic temporal evolution of \RSi,
starting from a high value well before maximum and leveling-out just
before maximum, as was the case for SN~2002bo \citep{ben04}. On the
other hand, the LVG SNe show on average either no evolution in \RSi\
before maximum or an evolution in the opposite sense in the case of
SN~1990N.  Clearly, the number of SNe~Ia for which very early spectra
are available is still too small to draw definite conclusions, but it
is worth speculating on a possible interpretation.

Since \RSi\ is related to the photospheric temperature of the ejecta
\citep{nug95}, the curves in Fig. 2 should trace the temperature
evolution in the line forming regions before maximum light.  HVG SNe seem to
start at cooler temperatures, which then increase approaching maximum. LVG SNe,
on the other hand, have high temperatures already well before maximum.
This may be related to the fact that LVG SNe have lower expansion velocities,
especially before maximum, and are also hotter.

\subsection{\RSi$_{max}$ vs. \dm} \label{nug}

In Figure 3a, the value of \RSi$_{max}$ measured for each SN~Ia at maximum
light is plotted against \dm. SNe in both the FAINT and HVG groups seem to
follow the relation between \RSi$_{max}$ and \dm\ established by
\citet{nug95}. LVG SNe, on the other hand, are more scattered in this plot,
especially at the bright, slow end. Either they do not conform with the
\RSi$_{max}$ -- \dm\ relation or they define a new, looser one.

\begin{figure}[t]
\figurenum{3}
\includegraphics[width=8.4cm, angle=0]{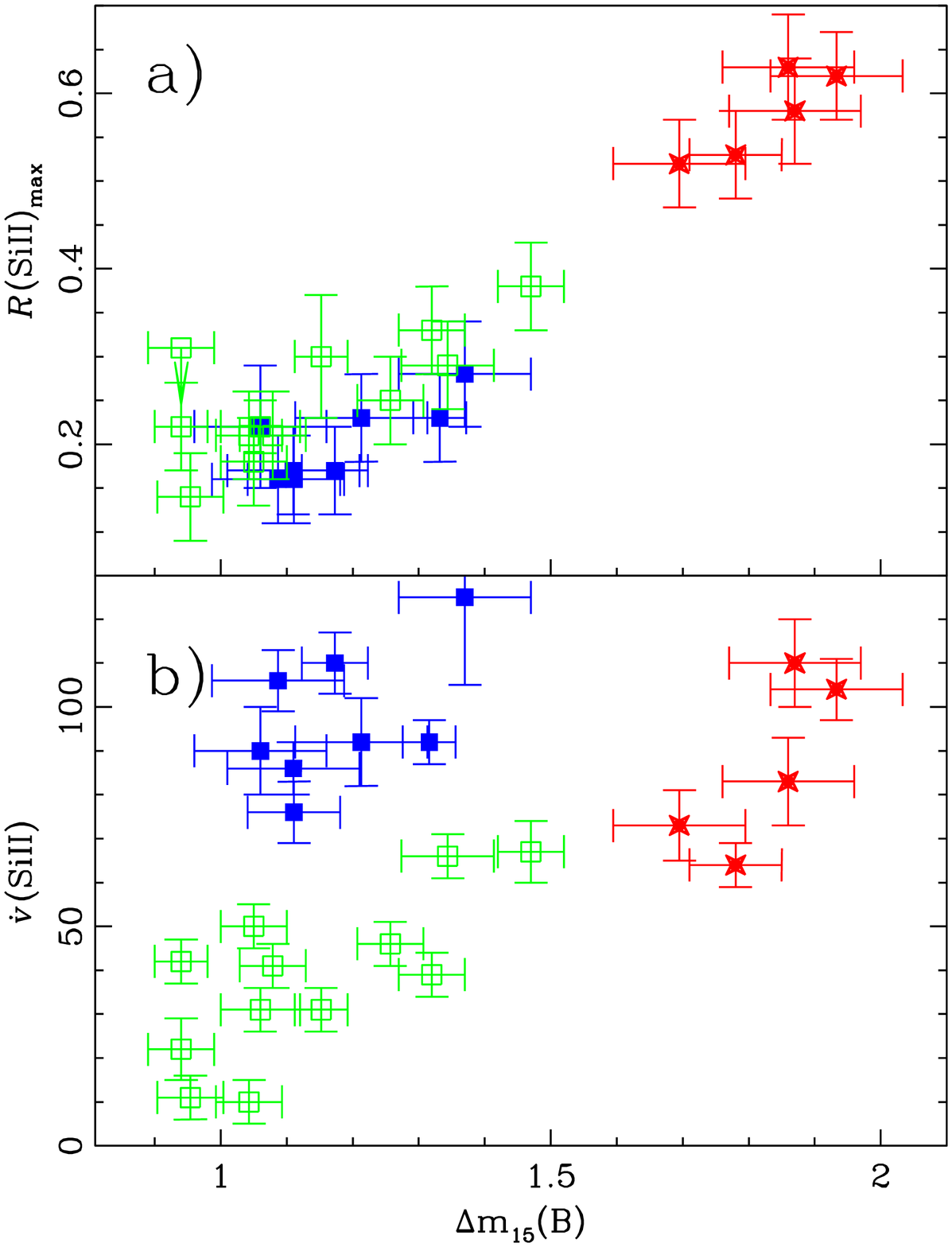}
\caption{a). \dm\ vs. \RSi$_{max}$\ for the SNe of our sample.  
b). \dm\ vs. $\dot{v}$ of SiII $\lambda 6355$. 
Open symbols refer to LVG SNe, filled symbols to HVG SNe, and starred symbols 
to FAINT SNe.
}
\end{figure}

The scatter of LVG SNe in Figure 3a, especially at low \dm, suggests
that another physical parameter besides the temperature is needed to
describe their behavior.  \citet{gar04}, using SYNOW synthetic
spectra, tentatively suggest that in SN~1999by and, in general, in
SNe~Ia with \dm$>1.2$ the 5800 \AA\ feature is mostly due to TiII
transitions rather than to SiII, and claim that only with this
interpretation can the \RSi$_{max}$ -- \dm\ relation be understood.\\
However, neither our synthetic spectral analysis of SN~1991bg
\citep{maz97}, nor the spectral tomography analysis of the normal
SN~Ia, SN~2002bo exploring a wide range of photospheric temperatures
and chemical compositions \citep{ste04}, requires a relevant contribution
from the TiII transitions to fit the 5800 \AA\ feature. Indeed, in the
above mentioned models, the 5800 \AA\ feature is always well fitted by
the SiII 5972 \AA\ transition.\\ 
As for the SNe~Ia with \dm$<1.2$, \cite{gar04} confirm that the 5800
\AA\ feature is indeed SiII.

\subsection{$\dot{v}$ vs. \dm}

The \dm\ parameter is plotted versus $\dot{v}$ in Figure~3b. The
expansion velocity evolution gradient, $\dot{v}$, seems to be weakly
correlated with \dm: while SNe with a large \dm\ (the FAINT group)
have a large $\dot{v}$, normal SNe can have both a large or a small
$\dot{v}$. Cluster analysis however suggests that we are dealing with
three distinct families of SNe~Ia: FAINT, LVG, and HVG.  These three
groups may be characterized by different physical parameters governing
the same explosion mechanism (possibly distinguishing LVG and HVG) or
by a totally different kind of explosion, which may be the case
especially for the FAINT group.

\section{DISCUSSION} \label{dis}

Based on the evidence presented above, we can make a preliminary attempt to 
explain the causes of the spectroscopic diversity among SNe~Ia. 

As Figures 1,2 and 3 show, SNe~Ia can be divided into three groups. 
Each group has distinct physical properties, different from those of 
other groups. In particular, the FAINT group (essentially the
SN~1991bg-like objects, plus SNe~1986G and 1993H according to this
method - but see \citet{mei00} for an infrared view) clearly differs
from the other two: these SNe are fast decliners in both luminosity
and velocity, they have typically low velocities and occur in
earlier-type galaxies. This may be not surprising, since in many ways
SN~1991bg--like objects stand out as odd.

For the other two groups, the situation is more complicated: both HVG and LVG
include normal SNe, but the LVG include also all the brightest, slowest 
declining SNe. Our analysis suggests that LVG and HVG are two distinct groups,
but they may possibly represent a continuum of properties. 

Interestingly, although it is common to refer to two main groups of peculiar
SNe~Ia, SN~1991bg-- and SN~1991T--like, the latter SNe fall in the same class
as  normal SNe, while the former do not.  There appears to be a discontinuity
of properties between SN~1991bg-like objects and all other SNe~Ia, which is not
seen for SN~1991T--like SNe. This is not what we might expect if SNe~Ia behaved
as a simple one-parameter family of events. 

Based on a qualitative analysis \citet{ben04} suggested that the large
blueshift of the SiII line in SN~2002bo (a HVG SN) maybe the result of
a delayed detonation explosion. Moreover, \cite{len00} find, from
detailed non-LTE calculations, that some delayed-detonation models can
account for the very high SiII blueshift of another HVG, SN~1984A. It
may be that the HVG SNe are delayed-detonations, their internal
dispersion arising from a range of transition densities, while the LVG
SNe are deflagrations. This would be an extension of the results of
\citet{hat00} (see their Figure 1).  The only difference with their
conclusions would be that SNe~1981B and 1992A would become a
delayed-detonation and a deflagration event, respectively. From
spectropolarimetric studies \citet{wan04} also reached the conclusion
that the explosion mechanism of SNe~1984A, 1997bp, 2002bo (all HVG SNe) 
and of SN~2004dt, which is most probably also a HVG SN, may be markedly 
different from that of lower velocity objects such as SN~1994D.

The LVG include all three SNe in our sample with \dm$ < 1$:
SNe~1991T, 1999cw and 1999ee. SN~1991T has also often been discussed
as the result of a delayed detonation, especially in order to explain
the high abundance of $^{56}$Ni and its decay products at the highest
velocities \citep{maz95}, so its inclusion in the LVG does not support
the hypothesis that all LVG are deflagrations. Very early measurements
of $v$(Si) are not possible for SN~1991T, since the SiII line was almost 
absent in the earliest spectra owing to the high degree of ionization.
The photospheric velocities inferred from spectral models were however 
very high \citep{maz95}. This might support the suggestion made by 
\citet{wan04} that SN~1991T-like events could be 1984A-like events 
viewed at different angles (and thus HVGs). 

Differences in the properties of the outer ejecta, such as a different
degrees of mixing, or of circum-stellar interaction, may also be at
the origin of the difference between LVG and HVG.

More efficient mixing out of heavy elements might result in an
initially higher photospheric velocity. At the earliest times the
photosphere should in fact tend to trace the heavier elements, since
they have much larger line opacity and this is the major contributor
to the optical depth \citep{paul96}. The initial rate of decrease of
the velocity with time would consequently be larger, as the
photosphere moves inward to layers that are not so different
from those of less mixed SNe. This would also mean an initially
lower temperature (resulting from the large photosphere) but
increasing with time. This would be the HVG group.

On the other hand, less efficient mixing could lead to initially
smaller photospheric velocities: light elements contribute much less
to the opacity and thus the photosphere would be deeper and the
pre-maximum temperature higher. The decline rate of the velocity would
then be smaller, and the temperature would either stay constant or
decline, depending on the exact combination of increasing luminosity
and decreasing photospheric velocity. These are the properties of the
LVG.

Very high-velocity features have been observed in all SNe with sufficiently
early spectra (Mazzali et al., in preparation). Maybe in the LVG the
interaction affects the spectra only very early, as in SN~1999ee
\citep{maz05} or SN~1990N \citep{fish97,maz01}, and there is a sudden
drop to lower velocities when the interaction ends, resulting in a
lower post-maximum $\dot{v}$, while it continues for a longer time in
the HVG, so that \vSiiX\ spans a broader range of values.

In both of these last scenarios, LVG and HVG SNe would not necessarily be
differentiated by the nature of the explosion, and they may even represent a
continuum of properties.

\acknowledgments 
We are in debt to J. Sollerman, P. Lundqvist, M.M. Phillips, S. Mattila and J.C Wheeler for
providing us unpublished SNe spectra. We also acknowledge D.A. Howell
and L. Wang for sending us the spectra of SNe 1999by and 2001el
respectively. This work is supported in part by the European
Community's Human Potential Programme under contract
HPRN-CT-2002-00303, ``The Physics of Type Ia Supernovae''.


\begin{thebibliography}{}

\bibitem[Altavilla et al.(2004)]{alt04} Altavilla, G., et al.\
2004, \mnras, 349, 1344 (A04)

\bibitem[Anderberg (1973)]{and73} Anderberg, M.R., 1973, Cluster
Analysis for Applications. Academic Press, New York

\bibitem[Barbon, Rosino, \& Iijima(1989)]{bar89} Barbon, R., Rosino,
L., \& Iijima, T.\ 1989, \aap, 220, 83 (Ba89)

\bibitem[Barbon et al.(1990)]{bar90} Barbon, R., Benetti, S., 
Rosino, L., Cappellaro, E., \& Turatto, M.\ 1990, \aap, 237, 79 (B90)

\bibitem[Benetti (1989)]{ben89}
Benetti, S.\ 1989, Degree Thesis, Universit\'a di Padova (B89)

\bibitem[Benetti, Cappellaro, \& Turatto(1991)]{ben91} 
Benetti, S., Cappellaro, E., \& Turatto, M.\ 1991, \aap, 247, 410 (B91)

\bibitem[Benetti et al.(2004)]{ben04} Benetti, S., et al.\ 
2004a, \mnras, 348, 261 (B04)


\bibitem[Bonanos et al.(1999)]{bon99} Bonanos, A., Garnavich, 
P., Schlegel, E., Jha, S., Challis, P., Kirshner, R., Hatano, K., \& 
Branch, D.\ 1999, Bull. Am. Astron. Soc., 31, 1424 (B99)

\bibitem[Branch et al.(1983)]{bra83} Branch, D., Lacy, C.~H., 
McCall, M.~L., Sutherland, P.~G., Uomoto, A., Wheeler, J.~C., \& Wills, 
B.~J.\ 1983, \apj, 270, 123 (B83)

\bibitem[Branch(1987)]{bra87} Branch, D.\ 1987, \apjl, 316, 
L81 

\bibitem[Branch \& van den Bergh(1993)]{bra93a} Branch, D.~\& van den
Bergh, S.\ 1993, \aj, 105, 2231

\bibitem[Branch, Fisher, \& Nugent(1993)]{bra93b} Branch, D., 
Fisher, A., \& Nugent, P.\ 1993, \aj, 106, 2383 

%
\bibitem[Buta, Corwin, \& Opal(1985)]{but85} Buta, R.J., 
Corwin, H.G., \& Opal, C.B.\ 1985, \pasp, 97, 229 

\bibitem[Cristiani et al.(1992)]{cri92} Cristiani, S., et 
al.\ 1992, \aap, 259, 63 (C92)

\bibitem[Fisher et al.(1997)]{fish97} Fisher, A., Branch, D., Nugent, P., 
Baron, E. 1997, \apj, 481, L89 

\bibitem[Freedman et al.(2001)]{fre01} Freedman, W.~L., et 
al.\ 2001, \apj, 553, 47 

\bibitem[Garnavich et al. (2004)]{gar04} Garnavich, P.M., et al.\ 2004,
\apj, in press (astro-ph/0105490)

\bibitem[Hamuy et al.(2002)]{ham02} Hamuy, M., et al.\ 2002, \aj, 124,
2339 (H02)

\bibitem[Harris et al.(1983)]{har83} Harris, G.L., Hesser, J.E., 
Massey, P., Peterson, C.J., \& Yamanaka, J.M.\ 1983, \pasp, 95, 607 

\bibitem[Hatano et al.(2000)]{hat00} Hatano, K., Branch, D., Lentz, E.J., 
Baron, E., Filippenko, A.V., \& Garnavich, P.M.\ 2000, \apjl, 543, L49 

\bibitem[Hernandez et al.(2000)]{her00} Hernandez, M., et 
al.\ 2000, \mnras, 319, 223 (H00)


\bibitem[Howell, H{\" o}flich, Wang, \& Wheeler(2001)]{how01} 
Howell, D.A., H{\" o}flich, P., Wang, L., \& Wheeler, J.C.\ 2001, \apj, 
556, 302 (H01)
 
\bibitem[Kirshner et al.(1993)]{kir93} Kirshner, R.P., et 
al.\ 1993, \apj, 415, 589 (K93)

\bibitem[Kraan-Korteweg(1982)]{kra82} Kraan-Korteweg, R.C.\ 
1986, \aaps, 66, 255 

\bibitem[Krisciunas et al.(2003)]{kri03} Krisciunas, K., et 
al.\ 2003, \aj, 125, 166 (K03)

\bibitem[Leibundgut et al.(1991)]{lei91} Leibundgut, B., Kirshner, R.P., 
Filippenko, A.V., Shields, J.C., Foltz, C.B., Phillips, M.M., \& 
Sonneborn, G.\ 1991, \apjl, 371, L23 (L91)

\bibitem[Lentz et al.(2000)]{len00} Lentz, E.J., Baron, E., 
Branch, D., Hauschildt, P.H., \& Nugent, P.E.\ 2000, \apj, 530, 966 

\bibitem[Li et al.(1999)]{li99} Li, W.~D., et al.\ 1999, 
\aj, 117, 2709 (L99)

\bibitem[Li et al.(2001)]{li01} Li, W., et al.\ 2001, \pasp, 
113, 1178 

\bibitem[Li et al.(2003)]{li03} Li, W., et al.\ 2003, \pasp, 
115, 453

\bibitem[Mazzali et al.(1997)]{maz97} Mazzali, P.~A., Chugai, 
N., Turatto, M., Lucy, L.~B., Danziger, I.~J., Cappellaro, E., della Valle, 
M., \& Benetti, S.\ 1997, \mnras, 284, 151 

\bibitem[Mazzali(2001)]{maz01} Mazzali, P.~A.\ 2001, \mnras, 321, 341

\bibitem[Mazzali, Danziger, \& Turatto(1995)]{maz95} Mazzali, P.A., 
Danziger, I.J., \& Turatto, M.\ 1995, \aap, 297, 509 (M95)

\bibitem[Mazzali et al.(2005)]{maz05} Mazzali, P.~A., Benetti, S., Stehle, M.,
Branch, D., Deng, J., Maeda, K., Nomoto, K., \& Hamuy, M. 2005, \mnras,
submitted
%
\bibitem[Meikle(2000)]{mei00} Meikle, W.P.S.\ 2000, \mnras, 314, 782 

\bibitem[Nugent et al.(1995)]{nug95} Nugent, P., Phillips, 
M., Baron, E., Branch, D., \& Hauschildt, P.\ 1995, \apjl, 455, L147

\bibitem[Patat et al.(1996)]{pat96} Patat, F., Benetti, S., 
Cappellaro, E., Danziger, I.J., della Valle, M., Mazzali, P.A., \& 
Turatto, M.\ 1996, \mnras, 278, 111 (P96)

\bibitem[Pauldrach et al.(1996)]{paul96} Pauldrach, A.W.A., Duschinger, M., 
Mazzali, P.A., Puls, J., Lennon, M., Miller, D.L.\ 1996 \aap, 312, 525 

\bibitem[Phillips et al.(1987)]{phi87} Phillips, M.M., et 
al.\ 1987, \pasp, 99, 592 (P87)

\bibitem[Phillips et al.(1992)]{phi92} Phillips, M.M., Wells, L.A., 
Suntzeff, N.B., Hamuy, M., Leibundgut, B., Kirshner, R.P., 
\& Foltz, C.B.\ 1992, \aj, 103, 1632 (P92)

\bibitem[Phillips(1993)]{phi93} Phillips, M.~M.\ 1993, \apjl, 
413, L105 

\bibitem[Phillips et al.(1999)]{phil99} Phillips, M.M., Lira, 
P., Suntzeff, N.B., Schommer, R.A., Hamuy, M., \& Maza, J.\ 1999, \aj, 
118, 1766 (P99)

\bibitem[Pignata et al.(2004)]{pig04} Pignata, S., et al.\ 
2004, \mnras, in press (P04)

\bibitem[Salvo et al.(2001)]{sal01} Salvo, M.E., Cappellaro, E., 
Mazzali, P.A., Benetti, S., Danziger, I.J., Patat, F., \& Turatto, M.\ 
2001, \mnras, 321, 254 (S01)

\bibitem[Stehle et al.(2004)]{ste04} Stehle, M., Mazzali, P.A., Benetti, S.,
Hillebrandt, W. 2004, \mnras, submited (astro-ph/0409342)

\bibitem[Stritzinger et al.(2002)]{stri02} Stritzinger, M., et al.\ 2002, 
\aj, 124, 2100 (S02)

\bibitem[Tsvetkov(1985)]{tsv85} Tsvetkov, D.Y.\ 1985, Soviet 
Astronomy, 29, 211 


\bibitem[Turatto et al.(1998)]{tur98} Turatto, M., Piemonte, A., 
Benetti, S., Cappellaro, E., Mazzali, P.A., Danziger, I.J., \& Patat, F.\ 
1998, \aj, 116, 2431 (T98)

\bibitem[Vink{\' o} et al.(2001)]{vin01} Vink{\' o}, J., 
Kiss, L.L., Cs{\' a}k, B., F{\H u}r{\' e}sz, G., Szab{\' o}, R., Thomson, 
J.R., \& Mochnacki, S.W.\ 2001, \aj, 121, 3127 (V01)

\bibitem[Wang et al.(2003)]{wan03} Wang, L., et al.\ 2003, 
\apj, 591, 1110 (W03)

\bibitem[Wang et al.(2004)]{wan04} Wang, L, Baade, D., Hoeflich, P., Wheeler, J.C., Kawabata, K., Khokhlov, A.,
Nomoto, K., Patat, F., 2004, ApJ, submitted, astro-ph/0409593.

\bibitem[Wegner \& McMahan(1987)]{weg87} Wegner, G.~\& 
McMahan, R.~K.\ 1987, \aj, 93, 287 (W87)

\bibitem[Wells et al.(1994)]{wel94} Wells, L.~A., et al.\ 
1994, \aj, 108, 2233 (W94)

\end{thebibliography}
\end{document}